\documentclass[12pt]{iopart}
\usepackage{amssymb}
\usepackage{epsfig} 
\usepackage{graphicx} 
\usepackage{epstopdf}
\usepackage{float}
\usepackage{hyperref} 
\def\beq {\begin{equation}}
\def\eeq {\end{equation}}
\def\bea {\begin{eqnarray}}
\def\eea {\end{eqnarray}}
\def \PMET{p{\!\!\!/}_T}
\begin{document}

\title{Searching the sbottom in the four lepton channel at the LHC}

\author{Diptimoy Ghosh$^a$, Dipan Sengupta$^b$}
\address{$^a$ Department of Theoretical Physics, Tata Institute of 
Fundamental Research, Homi Bhabha Road, Mumbai, India}
\ead{diptimoyghosh@theory.tifr.res.in}

\address{$^b$ Department of High Energy Physics, Tata Institute of Fundamental 
Research,Homi Bhabha Road, Mumbai, India}
\ead{dipan@tifr.res.in}


\begin{abstract} 

Direct searches at the Large Hadron Collider (LHC) have pushed the lower 
limits on the masses of the gluinos ($\tilde{g}$) and the squarks of the 
first two generations ($\tilde{q}$) to the TeV range. On the other hand, 
the limits are rather weak for the third generation squarks and masses 
around a few hundred GeV are still allowed. A comparatively light third 
generation of squarks is also consistent with the lightest Higgs boson 
with mass $\sim$ 125 GeV. In view of this, we consider the direct 
production of a pair of sbottom quarks ($\tilde{b}_1$) at the LHC and 
study their collider signatures. We focus on the scenario where the 
$\tilde{b}_1$ is not the next-to-lightest supersymmetric particle (NLSP) 
and hence can also decay to channels other than the commonly considered 
decay mode to a bottom quark and the lightest neutralino 
($\tilde{\chi}^0_1$). For example, we consider the decay modes 
containing a bottom quark and the second neutralino ($\tilde{b}_1 \to b 
\tilde{\chi}^0_2$) and/or a top quark and the lightest chargino 
($\tilde{b}_1 \to t \tilde{\chi}^{\pm}_1$) following the leptonic decays 
of the neutralino, chargino and the top quark giving rise to a 4 leptons 
($\ell$) + 2 $b$-jets + missing transverse momentum ($\PMET$) final state. 
We show that an sbottom mass $\lesssim$ 550 GeV can be probed in this channel
 at the 14 TeV LHC energy with integrated luminosity $\lesssim$ 100 fb$^{-1}$.

\end{abstract}

\maketitle
\section{Introduction}
\label{intro}

The Large Hadron Collider (LHC) is currently operational at the center 
of mass (c.m.) energy of 8 TeV and both the experiments CMS and ATLAS have 
collected about 10 fb$^{-1}$ of data each. The hint of a Standard Model (SM)
 like Higgs boson with mass 
around 125 GeV has been reported 
\cite{:2012gu,:2012gk,ATLAS-CONF-2012-103,ATLAS-CONF-2012-109}
 and it has spurred a large number 
of studies specially in the context of constraining 
and probing physics beyond the standard model (BSM). Supersymmetry 
(SUSY) has been a leading candidate for BSM for more than three decades and 
constitutes a major search program at the LHC. As of now a huge amount 
of data has been analyzed in the context of the Minimal Supersymmetric 
Standard Model (MSSM) and limits have been placed on the MSSM parameter 
space particularly in the framework of constrained MSSM (cMSSM). 
The current limits on squarks and gluinos from direct 
searches stand around 
1.5 TeV for approximately equal squark and gluino masses and about 950 
GeV for the case where gluino masses are much smaller than the squark 
masses \cite{CMS-PAS-SUS-12-005} The discovery of a light SM like Higgs boson 
however has put SUSY and in particular the cMSSM into perspective. A 
large number of papers have been written in this context 
\cite{Baer:2011ab,Akula:2011aa,Feng:2011aa,Heinemeyer:2011aa, Buchmueller:2011ab,
Draper:2011aa,Cao:2011sn,Hall:2011aa,Ellis:2012aa,Cao:2012fz,Chang:2012gp,
Baer:2012uya,Ghosh:2012mc,Maiani:2012ij,Cheng:2012np,Jegerlehner:2012ju, 
Choudhury:2012tc,Byakti:2012qk,Brummer:2012ns,Balazs:2012qc,Badziak:2012mm,
Feng:2012jf,Ghosh:2012dh,Dudas:2012hx,Fowlie:2012im,Athron:2012sq,
Chatterjee:2012qt,CahillRowley:2012rv,Akula:2012kk,Cao:2012yn,
Arbey:2012dq,Nath:2012fa} and the message from them seems to 
suggest that the SUSY parameter space is now extremely constrained, 
specially the gluinos and the first two generations of squarks.

The third generation of squarks is however special owing to the large 
Yukawa couplings and hence can decouple from the first two generations 
of squarks to become comparatively lighter. Light stop squarks 
($\tilde{t}$) are also favorable in order to cancel large radiative 
corrections from the top quark in Higgs mass and hence a necessity to 
reduce the problem of fine tuning in the SM.
 The signatures for a light 
stop quark at the LHC have been studied extensively in the past and have 
recently seen a flurry of activities \cite{Desai:2011th,
He:2011tp,Drees:2012dd,Berger:2012ec,Plehn:2012pr,Han:2012fw,Barger:2012hr,
Choudhury:2012kn,Cao:2012rz,Ghosh:2012ud,Dutta:2012kx,Chen:2012uw}.
As $\tilde{t_L}$ and $\tilde{b_L}$ belong to the same weak doublet, 
a light stop mass eigenstate 
may also be associated with a light sbottom mass eigenstate in specific 
scenarios.

In this paper we consider the possibility of a light third generation of 
squarks, in particular we focus on a light sbottom and investigate the 
viability of its signal at the LHC with 14 TeV c.m. energy. We do not 
confine ourselves to a particular SUSY breaking scenario and perform our 
study without assuming any relations among the soft SUSY breaking 
parameters at the electroweak scale.

Studies on the prospect of an sbottom search at the LHC, although not 
neglected in literature, are rather sparse. Some of the earliest studies
of sbottom phenomenology at colliders were performed in 
\cite{Bartl:1997yi,Hisano:2003qu,Drees:2000he,Hisano:2002xq}.
A study on the possibility of determining the sbottom spin at the LHC using 
angular correlations was performed in \cite{Alves:2007xt}. 
It should be noted that the 
sbottom pair production cross section is at par with that of stop and 
hence sbottom search should be conducted with the same priority as stop 
searches. In fact, differing topologies in various scenarios (leptons, 
b-jets etc.) can be used to distinguish between stop and sbottom and can 
provide useful information about the nature of SUSY parameter space in 
question. Hence sbottom search at the LHC can be complementary to stop 
quark searches. Study of the prospect of a SUSY signal in a scenario where the 
sbottom is the NLSP has been performed in the literature in the channel 
$\tilde{b}_{1}\to b \tilde{\chi}_{1}^{0}$ in the context of both LHC and ILC \cite{Belyaev:2009wf,AdeelAjaib:2011ec,Lee:2012sy,Alvarez:2012wf,Ajaib:2012eb}.

Recently the CMS collaboration ruled out sbottom mass up to 500 GeV 
with 4.98 fb$^{-1}$ of 7 TeV data assuming the branching ratio 
${\rm BR}(\tilde{b}_{1}\to b \tilde{\chi}_{1}^{0})$ to be 100\%
and the LSP mass of about 175 GeV \cite{CMS:2012nxa}. 
This exclusion was also crucially dependent on the LSP mass and 
there was no exclusion limit for the LSP mass of about 200 GeV or higher.

However, in a large part of the MSSM parameter 
space the sbottom is not the NLSP. As a consequence, the branching ratios 
(BR) to channels other than $\tilde{b}_1 \to b \tilde{\chi}^0_1$ may be 
significant. 
Recently both the ATLAS and CMS collaborations have also searched for 
sbottoms in the leptonic channel in the decay mode $\tilde{b}_{1}\to t 
\tilde{\chi_{1}}^{\pm}$ and in the hadronic mode with b-tagged jets in the 
$\tilde{b}_{1}\to b \tilde{\chi}_{1}^{0}$ channel and have constrained a 
narrow region of parameter space assuming specific mass relations among 
$\tilde{b}_{1}$, $\tilde{\chi}_{1}^{0}$ and 
$\tilde{\chi}_{1}^{\pm}$ \cite{Chatrchyan:2012sa,Aad:2012pq}.
 For the leptonic channel the
$\tilde{b}_{1}$ exclusion limits are $\sim$ 360-370 GeV for a
$\tilde{\chi}_{1}^{\pm}$ mass $\sim$ 180-190 GeV, and a
$\tilde{\chi}_{1}^{0}$ mass of 50 GeV. For the hadronic mode 
the exclusion limits are given in a  model
with gluino decaying into sbottom pairs  with further decay
into b-jets and lightest neutralino. The 
search excludes gluino masses around $\rm 1.1~TeV$ for sbottom masses in the range 
$\rm \sim 400-800$ GeV and a $\tilde{\chi}_{1}^{0}$ mass of 60 GeV.

\begin{figure}[h!]
\begin{tabular}{c}
\includegraphics[scale=0.4]{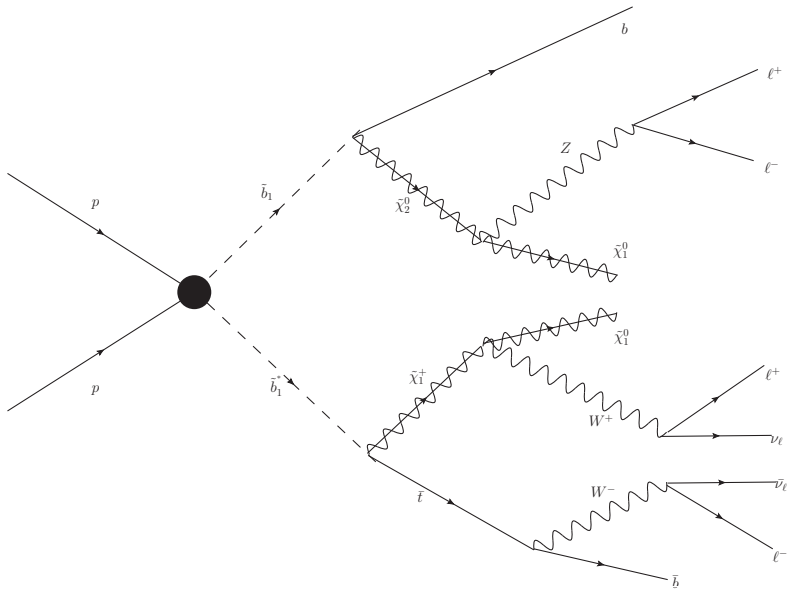}
\end{tabular}
\caption[]{ A sample Feynman diagram for the process 
$pp \to b \, \bar{b} \, \ell^+ \ell^+ \ell^- \ell^- + \PMET$ in MSSM.} 
\label{fd}
\end{figure}

In this paper we consider the decay of sbottom to the channels 
$\tilde{b}_{1}\to b \tilde{\chi}_{2}^{0}$ and $\tilde{b}_{1}\to t 
\tilde{\chi_{1}}^{\pm}$. The subsequent decays  of $\tilde{\chi_{2}}^{0}$ $\to$ 
$\tilde{\chi_{1}}^{0} Z$ and $\tilde{\chi_{1}}^{\pm}$ $\to$ 
$\tilde{\chi_{1}}^{0} W$ can now produce a number of hard leptons in 
the final state. A sample Feynman diagram is shown in Fig. \ref{fd}. 

%

The merits of considering the leptonic final state in 
particular, the 4 lepton channel is that it is rather clean and has 
minimal background. As we shall demonstrate below, it is possible to 
discover a SUSY signal for a substantial range of sbottom mass at 14 TeV LHC.

In the following section we compute the BR of sbottom to the above mentioned 
channels and choose some benchmark points for the collider study.  
We discuss the signal and backgrounds in section~\ref{SB}. Finally, we summarize 
our findings and conclude in section~\ref{concl}.

\section{SUSY Framework and Sbottom Branching Ratios}
\label{sec2}

If no specific mechanism for the SUSY breaking is assumed then the total 
number of unknown parameters (the so called soft SUSY breaking terms) 
reaches a huge number (105) and it is almost impossible to carry out any 
phenomenological analysis with such a large number of free parameters. 
Many of these parameters in particular the intergenerational mixing 
terms and the complex phases are rather constrained from various 
measurements of both Charge-Parity (CP) conserving and CP-violating 
observables in $K, B$ and $D$ decays as well as lepton flavor violating 
decays \cite{Martin:1997ns}.
 It is then phenomenologically useful to make a few 
assumptions (which are indeed supported by experiments) like no new 
source of CP violation, diagonal sfermion mass matrices and tri-linear 
couplings etc. to reduce the number of free parameters. This rather 
simplified version of MSSM is called a phenomenological MSSM (pMSSM) 
\cite{Djouadi:1998di}which has 22 free parameters. These parameters include

\begin{itemize} 

\item The gaugino (bino, wino and gluino) mass parameters $M_{1}$, 
$M_{2}$ and $M_{3}$.

\item The Higgs mass parameters $m_{H_u}$, $m_{H_d}$( which can be traded as 
$\mu$ and $M_{A}$)
 and the ratio of  the Vacuum Expectation Values (VEV) of the two Higgs
 doublet namely $\tan\beta$.

\item Common first and second generation sfermion mass parameters 
$m_{\tilde{Q}}, m_{\tilde{U}}, m_{\tilde{D}}, m_{\tilde{L}}, 
m_{\tilde{E}}$ and the third generation sfermion mass parameters 
$m_{\tilde{Q_3}}, m_{\tilde{t_R}}, m_{\tilde{b_R}}, m_{\tilde{L_3}}, 
m_{\tilde{\tau_R}}$.

\item The common first and second generation tri-linear couplings 
$A_{u}$, $A_{d}$ and $A_{e}$. The third generation tri-linear couplings 
$A_{t}$, $A_{b}$ and $A_{\tau}$.

\end{itemize}

In this work we take the pMSSM as our model framework and consider the 
constraints on the parameters coming only from the LEP exclusion limits
\cite{PhysRevD.86.010001}, theoretical considerations like correct electroweak symmetry 
breaking, electric and color neutral LSP etc. and the lightest Higgs 
mass in the range 123 -128 GeV. The 
tree level Higgs mass in MSSM is always less than the $Z$ boson mass and 
 is given by the formula $m_{h} {\bf \le} m_{Z}\cos(2\beta)$. Hence,
 large corrections from the virtual particles in the loop are required for the 
Higgs mass to be consistent with the range given above. The one loop 
contribution is dominated by the contribution from the stop and top 
quark sector and is given by \cite{Carena:2012gp}
\begin{equation}
\Delta m_{h}^{2} \simeq \frac{3}{4\pi^{2}} \frac{{m_{t}}^{4}}{v^2 
} \Bigg[ \frac{\tilde{X}_{t}}{2}+t + \frac{1}{16\pi^{2}}\Bigg(\frac{3}{2}\frac{m_{t}^{2}}{v^{2}}-32\pi\alpha_{3}\Bigg)\Bigg (\tilde{X}_{t}t+t^{2}\Bigg)\Bigg]
\label{stoploop} 
\end{equation}
where $ t=\textnormal{log}\frac{{M_{S}^{2}}}{{m_{t}^{2}}}$, 
$\tilde{X}_{t}=\frac{2\tilde{A}_{t}^{2}}{M_{s}^{2}}\Bigg(1-\frac{\tilde{A}_{t}^{2}}{12 M_{s}^{2}}\Bigg) $, 
$v$ =246 GeV, $ M_{S} = \sqrt{m_{\tilde{t}_{1}} 
m_{\tilde{t}_{2}}}$, the geometric average of the two stop masses and 
$\tilde{A_{t}}= A_{t} - \mu \cot\beta$, the mixing parameter in the stop sector.
If both the stop quark mass eigenstates are light, it will be difficult
to raise the Higgs mass to 125 GeV. However, with a large spliting
in the stop sector facilitated by a large $A_{t}$ term, it is possible 
to achieve one light eigenstate while the other heavy and thus achieving
the required Higgs mass. In this case the overall stop mass scale
 $ M_{s}=\sqrt{m_{\tilde{t}_{1}}m_{\tilde{t}_{2}}}$ can be of order TeV and 
hence the fine-tuning in the cancellation of the quadratic divergence 
will be small. On the other hand, this would require a large negetive 
value of $A_{t}$ introducing an adjustment of parameter in the model.

There can be important loop corrections from the bottom squark sector 
also and is given by \cite{Carena:2012gp}
\begin{equation}
\Delta m_{h}^{2}\simeq - \frac{h_{b}^{4}v^{2}}{16\pi^{2}}\frac{\mu^{4}}{M_{S}^{4}}(1 + \frac{t}{16\pi^{2}}(9h_{b}^{2}- 5\frac{m_{t}^{2}}{v^{2}}-64\pi\alpha_{3})).
\end{equation}
Here $h_{b}$ is the bottom Yukawa coupling which is given by 
$
h_{b}\simeq \frac{m_b}{v \cos\beta (1+\tan\beta \Delta h_b)},
$
where $\Delta h_b$ denotes the one loop correction\cite{Carena:2012gp}.

Hence, when $\mu \tan\beta$ is large the contribution from the sbottom 
sector can be non-negligible. Thus, probing the sbottom sector is also 
crucial to understand the Higgs sector and in turn the Electroweak Symmetry Breaking in the MSSM.

\begin{figure}[h!]
\begin{tabular}{c}
\includegraphics[scale=0.7]{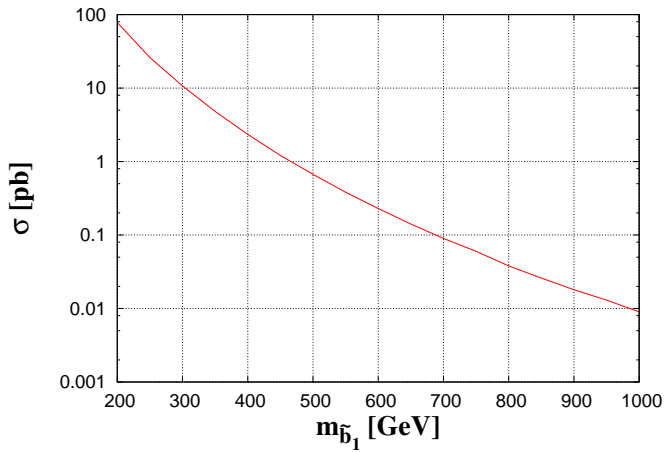}
\end{tabular}
\caption[]{ The central value of the Next to Leading Order (NLO) c.s. 
for the sbottom pair production at the 14 TeV LHC.} 
\label{cs}
\end{figure}

In Fig.~\ref{cs} we show the production c.s. of a sbottom pair at the 
14 TeV LHC. The cross sections are calculated using PROSPINO \cite{Beenakker:1996ed} 
in the limit where 1st two generation squarks are  
$\rm \sim 5~\rm TeV$, the gluino mass is around $\rm \sim 1.2~\rm TeV$ 
and the stop mass is around 400 GeV. The Renormalization and Factorizations 
scales are set to the their default values in PROSPINO and the CTEQ6L parton 
distribution function has been used for the c.s. calculation. 
It can be seen that the cross section falls sharply from 
$\rm \sim 10 pb$ at 300 GeV to $ \sim 10 \rm fb$ at
 $\rm 1\rm TeV$. It must be noted that the NLO cross section
 depends on the squark and gluino masses to some extent.
In our scenario the first two generation of
 squarks and the gluino is
decoupled from the 3rd generation squarks. 
  
%
%

The direct decay of sbottom to the LSP is always kinematicaly favored, 
and for right-handed squarks it can dominate if $\tilde{\chi}^0_1$ is 
bino like. This is generally the case with models like CMSSM 
which has a large right handed component in the third generation
mixing matrix. The bino co-annihilation case where the bino
can co-annihilate with the NLSP sbottom is also an important scenario
and has been considered in \cite{AdeelAjaib:2011ec}.  The interplay
of sbottom and stop in the context of natural SUSY has been discussed
in \cite{Lee:2012sy}, where the authors argue that direct searches
on sbottom can set limits on the stop sector from below casuing a
tension between naturalness which sets the stop scale from above
and direct searches which constrain it from below.They also suggest
that the limits on direct searches should depend on the admixture
of left and right handed components of sbottom. The left handed 
nature of sbottom has been searched by CMS in \cite{Chatrchyan:2012sa},
 for 7 TeV LHC where they investigate the channel 
$\tilde{b}_{1}\to t \tilde{\chi_{1}}^{\pm}$ in the dilepton + b-jets channel.
This motivates us to investigate the scope of MSSM to admit a large left handed
sbottom and ways to detect such a scenario.
If the sbottom is left handed then it may  
prefer to decay strongly into heavier charginos or neutralinos instead, for 
example $\tilde{b}_{1}\to b \tilde{\chi}_{2}^{0}$ and $\tilde{b}_{1}\to 
t \tilde{\chi_{1}}^{\pm}$. This is because the relevant 
squark-quark-wino couplings are much bigger than the squark-quark-bino 
couplings. Squark decays to higgsino-like charginos and neutralinos are 
less important for sbottom (than stop) because of its relatively smaller 
Yukawa coupling. A light left-handed sbottom can be achieved by a large 
splitting between the left-handed and the right-handed components 
 ($m_{\tilde{Q_3}}$ and $m_{\tilde{b_R}}$), in particular a light 
left-handed component ($m_{\tilde{{Q_3}}}$) and a heavy right-handed 
component ($m_{\tilde{b_R}}$). This ensures that once diagonalized the 
lighter sbottom remains predominantly left handed while the heavier 
sbottom remains mostly right handed.  The sbottom mixing matrix
in such a scenario is  diagonal with the mixing angle
$ \theta_{b}\sim 0$. For our purpose therefore, the 
relevant parameters are the third generation squark mass parameters 
($m_{\tilde{Q_3}}, m_{\tilde{t_R}}, m_{\tilde{b_R}}$), the tri-linear 
couplings $A_t$ and $A_b$, the SU(2) and U(1) gaugino mass parameters 
($M_1$ and $M_2$) and the Higgs sector parameters $\mu$ and $\tan\beta$.

To show that the situation we are considering is not a very fine tuned 
parameter space we vary the four parameters $m_{\tilde{Q_3}}, 
m_{\tilde{t_R}}, m_{\tilde{b_R}}$ in the range [100,3000] GeV and $A_t$ 
in the range [-3000, 3000] GeV and calculate the branching ratios
 of sbottom to different channels. We keep 
$\tan\beta=10$, $M_{1}=\rm 150 \, GeV$ and $M_{2}=250 \, \rm GeV$ in the scan.

 The first two generation squarks, and the three slepton generations
 are fixed at  $5 \, \rm TeV$ along with $M_{3}= 1 \rm TeV$, and 
$A_{u}=A_{d}=A_{\tau}=100 \, \rm GeV$ as they are irrelevant for our study.
The $\rm \mu$ parameter is set to 1000 GeV which implies that the 
lighter neutralino is gaugino like. We generate  the physical mass
spectrum using the spectrum generator SuSpect\cite{Djouadi:2002ze}.
A different set of choices for $\rm M_{1}$ and $\rm M_{2}$ do
not significantly alter the collider results significantly as long as 
$\chi_{2}^{0}\to\tilde{\chi}_{1}^{0} Z$ is kinematicaly allowed as can be 
observed in the next section.

\begin{figure}[h!]
\begin{tabular}{c}
\includegraphics[keepaspectratio=true,scale=0.6]{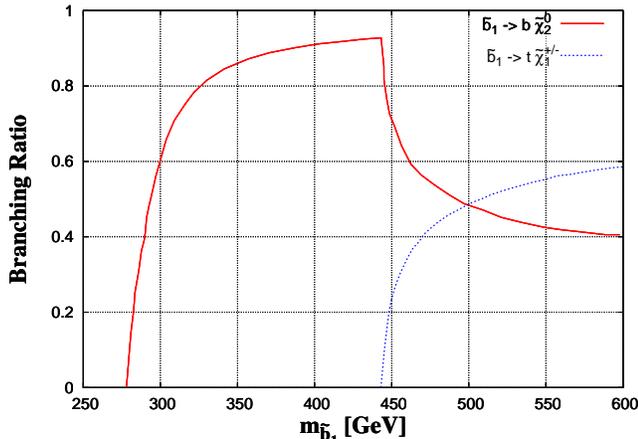}
\end{tabular}
\caption[]{ Maximum branching ratios of $\tilde{b}_{1}\to b\chi_{2}^{0}$ 
(red/continuous) and $\tilde{b}_{1}\to t\chi_{1}^{\pm}$ (blue/dotted) as 
a function of $\tilde{b}_{1}$ mass.} 
\label{br}
\end{figure}

 We choose $A_{b}=0$ GeV in our scan but other 
values do not change the result in a significant manner.
Fig \ref{br} shows the maximum values of the branching ratios 
for the channels $\tilde{b}_{1}\to b\chi_{2}^{0}$ and 
$\tilde{b}_{1}\to t\chi_{1}^{\pm}$ as a function of the sbottom mass when 
we vary the parameters in the ranges mentioned above. It can be seen that 
significant branching ratios to these channels are allowed.

\section{Signal and Background}
\label{SB}

In this section we choose a few benchmark points to carry out the 
collider analysis. The parameters for these benchmark points along with 
the relevant BRs are shown in Table.~\ref{tab1}. 

\begin{table}[h]
\begin{tabular}{|c|c|c|c|c|c|c|c|c|c|c|c|}
\hline
&  $A_{t}$  & $m_{\tilde{Q_3}}$  & $m_{\tilde{t}_{R}}$
& $m_{\tilde{b}_{R}}$  & $m_{\tilde{t}_{1}}$  & $m_{\tilde{b}_{1}}$  & 
$m_{\tilde{\chi}_1^{0}}$  & $m_{\tilde{\chi}_2^{0}} $  & 
$m_{\tilde{\chi}_1^{\pm}}$ & BR($\tilde{b}_{1}\to b\chi_{2}^{0}$) & 
BR($\tilde{b}_{1}\to t\chi_{1}^{\pm}$) \\
\hline
P1 & -2060 & 308 & 1922 & 1041 & 392 & 350  & 153 & 272  & 272 & 86 \% & --- \\
P2 & -2335 & 401 & 1907 & 2626 & 470 & 450 & 153 & 272 & 272 & 71\% & 24 \% \\
P3 & -2680 & 492 & 2232 & 1904 & 573.1 & 550 & 152 & 271 & 271 & 44.5\% & 54.5\%  \\
P4 & -2680 & 492 & 2232 & 1904 & 573.1 & 550 & 254 & 377 & 377 & 95\% & -   \\
\hline
\end{tabular} 
\caption{ Masses of some of the sparticles for three benchmark points.
In all the cases the other pMSSM parameters are fixed to values as described 
in the text. 
\label{tab1}}
\end{table}

In these parameter 
points the decay of sbottom proceeds mostly through the channels 
$\tilde{b}_{1}\to b \chi_{2}^{0}$ and/or $\tilde{b}_{1} \to t 
\chi_{1}^{\pm}$ following the deacys $\tilde{\chi}_{2}^{0} \to 
\tilde{\chi}_{1}^{0} Z \to l^{+} l^{-} \tilde{\chi}^{0}_{1}$ and 
$\tilde{\chi}_{1}^{\pm} \to \tilde{\chi}_{1}^{0} W^{\pm} \to l^{\pm} 
\nu_l \tilde{\chi}^{0}_{1}$ from both sides which finally yield a 
4-leptons + 2 b-jets + $\PMET$ signal in the final state.

A look at the spectrum and the decay branching ratios point out 
that in absence of a sufficient mass gap for the top decay to open up, 
the principal decay mode is $\tilde{b}_{1} \to b\tilde{\chi}_{2}^{0}  $. When 
the mass difference is sufficient for the top channel the 
branching ratio is fairly equally divided between the two channels.
This feature is also demonstrated in Fig \ref{br}.

As mentioned earlier, in order to show that a different choice of 
$M_{1}$ and $M_{2}$ do not change our results significantly as long as the decay $\tilde{\chi}_{2}^{0} \to \tilde{\chi}_{1}^{0} Z$ opens up we choose
the benchmark point P4 in Table \ref{tab1} in which $M_{1}$ and $ M_{2}$ are changed to 250 GeV (from 150 GeV in P1 - P3) and 350 GeV 
(from 250 GeV in P1 - P3) respectively.

We mentioned earlier that the signal cross sections falls sharply with 
increasing $\tilde{b}_{1}$ mass. In particular, for an sbottom of 550 GeV (P-3) 
the NLO cross section reduces to 385 fb which, because of the 
very low branching ratio for the leptonic decay modes of $Z$, yields a 
miniscule final cross section. However since the background is miniscule 
and we are optimistic about high luminosity options for a 14 TeV LHC, 
this channel still offers some hope even for such high sbottom mass.

In our simulation of events, we have used PYTHIA6 \cite{Sjostrand:2006za} for both the 
signal as well as the backgrounds. We construct jets using the FastJet 
\cite{Cacciari:2011ma} package employing the anti-$\rm k_{T}$ \cite{Cacciari:2008gp} 
algorithm with a cone size $\Delta R = 0.5$. We use the CTEQ6L \cite{Lai:1999wy} 
parton density function from the LHAPDF \cite{Bourilkov:2006cj} package. 
The scale is set at $Q^{2}=\hat{s}$. We then use the following selection criteria 
for the final events: 
\begin{enumerate}
\item We demand four isolated leptons (electron and muon) with the 
transverse momentum $p_{T}^{\ell} \ge 25$ GeV and the pseudo-rapidity 
$|\eta| \le 2.5$. Isolation of leptons are ensured by demanding the 
total transverse energy $p_T^{AC} \le 10\% $ of $p_T^{\ell}$. Here 
$p_T^{AC}$ is the scalar sum of transverse momenta of jets close to 
leptons satisfying $\Delta R(\ell,j) \le 0.2$ with a jet $p_{T}$ threshold 
of 30 GeV and $|\eta| \le 3$. We ensure that the sum total charge of the 
4 lepton system is 0 to avoid contamination from background.

\item Jets are selected with $p_{T}^{j} \ge 30$ GeV and $|\eta| \le 3$. 
We demand at least 2 jets with $b$ tags. 
The b-tagging is implemented by performing a matching of the jets 
with $b$ quarks assuming a matching cone $\Delta R(b,j) = 0.3$.

\item In addition we demand $\PMET > 50 $\,GeV.
\end{enumerate}  

The effects of the above selection cuts are summarized in Table.~\ref{tab2}. 
The potential SM backgrounds in our case are $t \bar{t}$, $Z\,Z$, $W\,Z$, 
and QCD. In Table.~\ref{tab2} we show only the non-vanishing background which 
in our case is the $t \bar{t}$.

\begin{table}[h] 
\begin{tabular}{|c|c|c|c|c|c|c|} 
\hline 
Process & Production c.s.  & Simulated Events & 4 isolated $\ell$& 
2 b-tagged jets &  
$\PMET$ $>$ 50 GeV & $\rm \frac{S}{\sqrt{B}}$ \\
 & (pb) & & & & & $\rm ({50 ~fb^{-1}})$\\
\hline \hline 
P1 & 4.75 & 0.5M  & 5.1 &  1.4 & 1.1  &  26 \\ 
P2 & 1.22 & 0.1M  & 1.5 &  0.5 & 0.4  &  10 \\
P3 & 0.39 & 0.1M  & 0.5 & 0.2  & 0.2  &  4 \\
P4 & 0.39 & 0.1M & 0.6 & 0.2 & 0.2  &  4  \\
\hline
$ \rm t \, \bar{t}$ & 918 & 40M & 2.8 & 0.09 & 0.09  \\
\cline{1-6} 
\end{tabular}
\caption{Efficiency of the selection cuts for the signal in the three 
benchmark points and the top background for 14 TeV LHC. The cross-sections after 
each of the cuts (column 4 - 6) are given in femtobarns. 
Efficiency for 2 $b$-tagging has been multiplied in the 5th column. 
The significance has been quoted at a projected luminosity of 
50 $\rm fb^{-1}$ in the last column.
\label{tab2}} 
\end{table} 

\begin{figure}[h!]
\begin{tabular}{cc}
\includegraphics[keepaspectratio=true,scale=0.4]{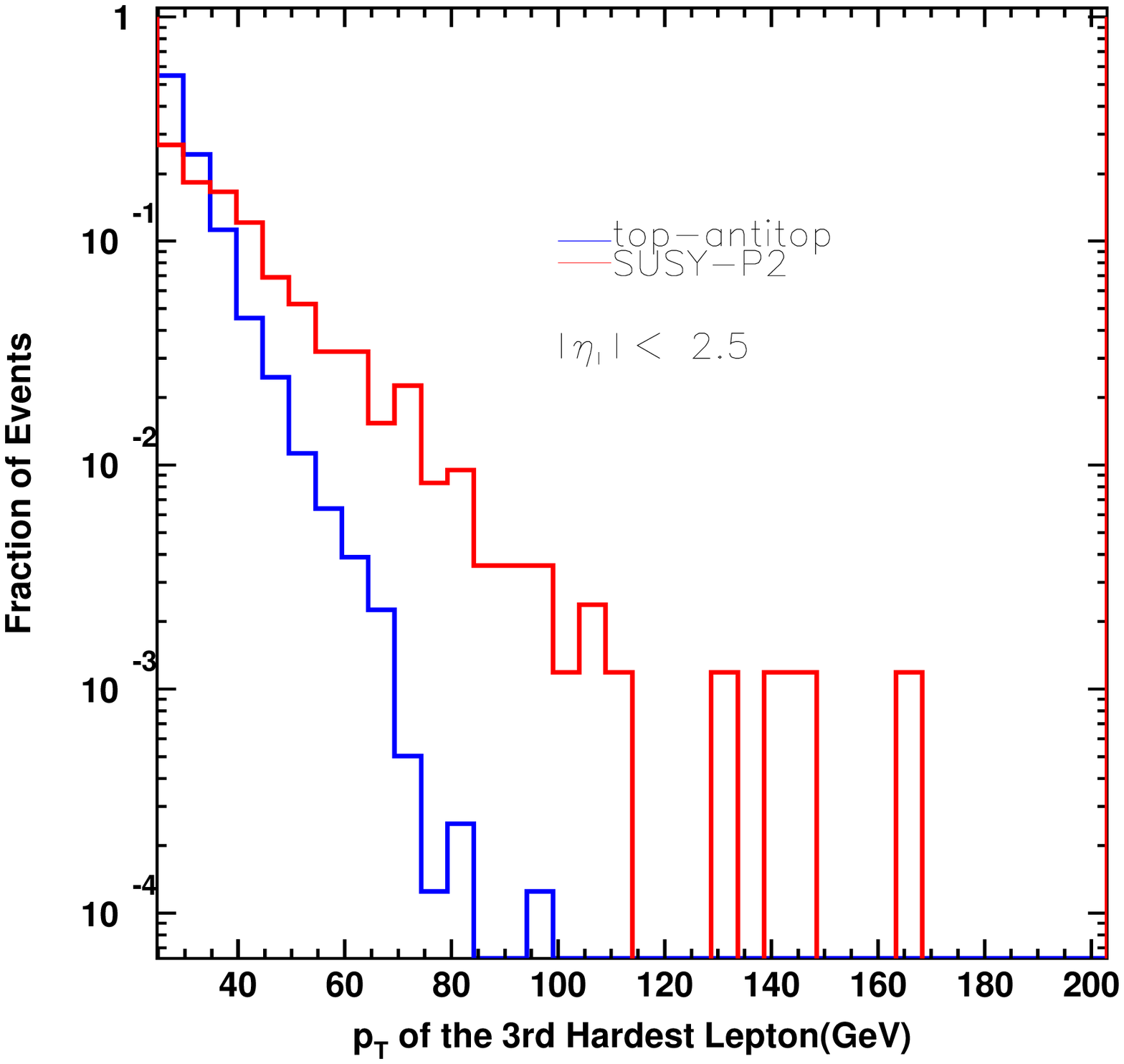}&
\includegraphics[keepaspectratio=true,scale=0.4]{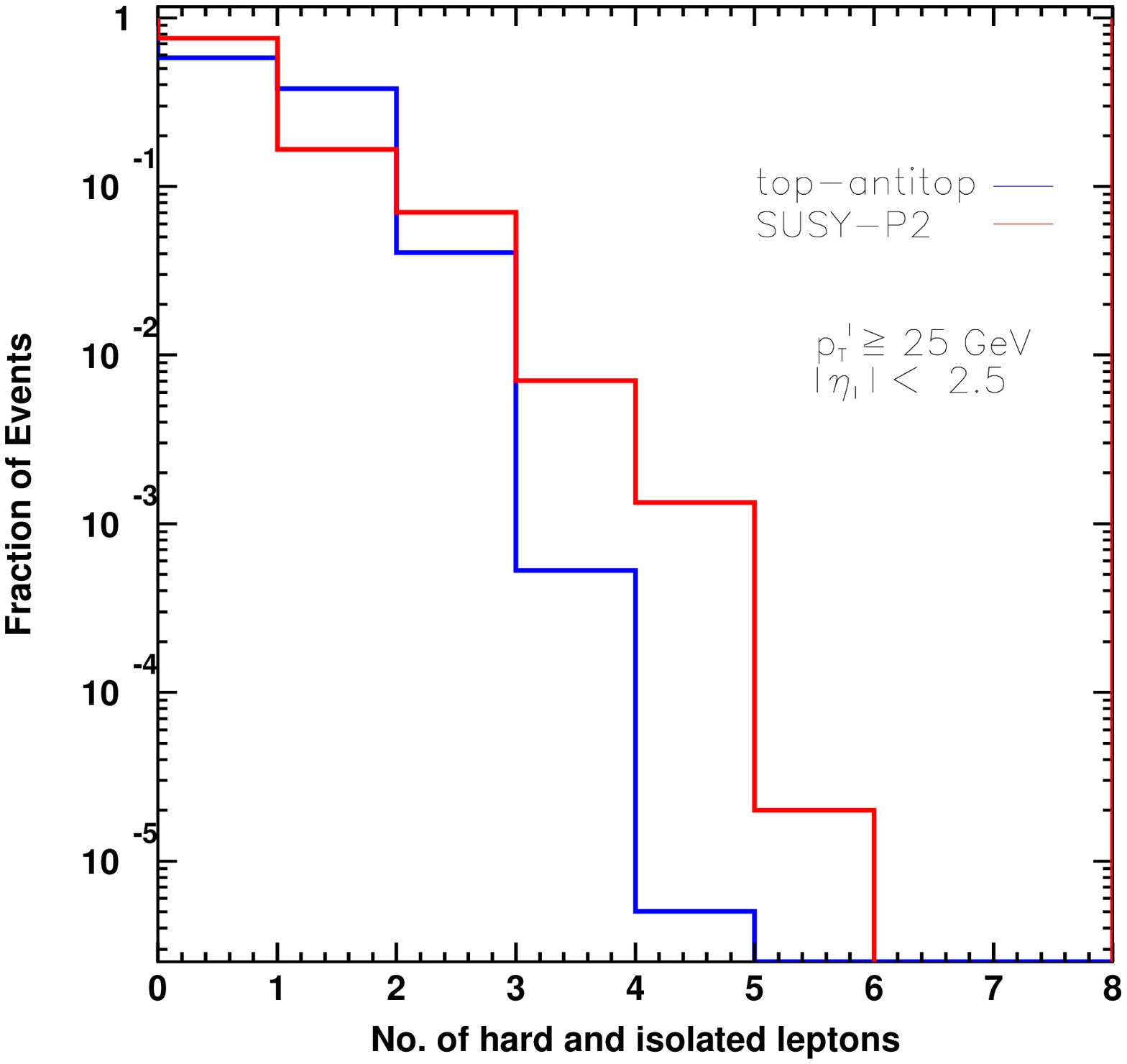}
\end{tabular}
\caption[]{ The $P_T$ distribution of the 3$^{\rm rd}$ hardest isolated (and $|\eta| < $ 2.5)
lepton (left panel) and the lepton 
multiplicities (right panel) for both the signal (benchmark-2) and the top 
background. 
\label{nlep}}
\end{figure}

The second column represents raw production c.s. for $\tilde{b}_{1} 
\bar{\tilde{b}}_{1}$ calculated at NLO using PROSPINO \cite{Beenakker:1996ed}. 
We have used the top pair production cross section at 14 TeV as quoted 
in Ref \cite{Kling:2012up}. The third column represents the number of events 
simulated for each of the processes. From the fourth column the cumulative effects of 
the selection cuts are shown. In demanding b-jets we assume an 
optimistic b-tagging efficiency of 70\% for each b-jet \cite{CMS-PAS-BTV-11-001}. We 
find that even for the signal the requirement of four isolated hard leptons 
with vanishing total lepton charge of the system leaves a small signal 
cross-section. On the other hand, this takes care of all the other backgrounds 
with the exception of $t \bar{t}$. The transverse momentum distribution of the 
3$^{\rm rd}$ hardest isolated (and $|\eta| < $ 2.5) lepton is shown in the left panel of 
Fig.\ref{nlep} where a clear distinction can be made between the signal and the 
background. The lepton multiplicities for both the
signal (benchmark-2) and $t \bar{t}$ background are also shown in the 
right panel of Fig.~\ref{nlep}. Note that, though in the parton level a  3$^{\rm rd}$ hard 
lepton is not expected from $t \bar{t}$ events, in real situation such leptons can 
come from the hadron decays for example, semileptonic decays of $B$ hadrons. 

The demand of two $b$-like jets in addition to the four isolated hard 
leptons also removes a significant fraction of the top background. 
Since the $\PMET$ is 
rather soft in the signal due to a low mass difference between 
$\tilde{\chi}_{2}^{0}$ and $\tilde{\chi}_{1}^{0}$, only a low $\PMET$ cut
could be used in selecting events. Our results are summmarized in the last column of 
Table.~\ref{tab2}. The signal significance is obtained in terms of 
Gaussian statistics, given by the ratio $S/\sqrt{B}$ of signal and 
background events for a particular integrated luminosity.
We project our significance ($S/\sqrt{B}$) at 50 fb$^{-1}$ at 14 TeV LHC.
We find that for low sbottom masses (up to 450 GeV) a reasonable 
significance ($S/\sqrt{B} \ge 5$) can be achieved at relatively low 
luminosities ($\sim$ 20 fb$^{-1}$). For masses of $\sim$ 500 GeV a higher 
luminosity of 50  fb$^{-1}$ will be required. For even higher masses the 
sbottom production c.s. is miniscule and it will require at least 
100 fb$^{-1}$ luminosity to get any hint of a signal at LHC.
As noted earlier we find that the change in LSP mass does not significantly change our signal significance. This can be seen in the event summary given 
in Table \ref{tab2}.

%
\section{Summary and Conclusion} 
\label{concl}
 We have probed the prospect of a light sbottom search in the 
 4 lepton + jets  (with two b-tagged jets) + $\PMET$ channel in the context of
pMSSM at 14 Tev LHC. We have considered the scenario where the lighter sbottom 
is predominantly left handed and can decay into the second lightest
neutralino or lighter chargino which eventually yields 4 leptons
and b jets. We find that in pMSSM there is a large part of parameter
space where such a scenario is feasible and can be useful to look for sbottom 
signatures. We also find that such a parameter space is compatible with a 
Higgs mass of 125 GeV and is in tune with the ongoing motivation for a 
light 3rd generation scenario.

We have analyzed the signal and background for such regions 
of parameter space and found that it is possible to discover
a substantial range of sbottom masses. In particular we find that
for sbottom masses $\sim 450$ GeV it is possible to find a viable signal at 
the level of $S/\sqrt{B}\ge 5$ even at 20 fb$^{-1}$ luminosity.
For masses $\sim$ 550 GeV and higher it will require higher luminosity LHC options which 
is achievable in the near future. It has to be noted that the channel has 
minimal background and the discovery reach is only cross section limited.
We have demonstrated that as long as the studied decay channel is kinematically
allowed our signal significance primarily depends on the signal cross section.
Hence in our study the LSP mass plays a less significant role as compared
to the NLSP sbotttom searches at the LHC which rely on a significant
mass splitting between the sbottom and the LSP. In order to show 
this we calculated the signal using two different values of LSP mass viz. 152 GeV and 254 GeV.

The 4-lepton channel in the context of 3rd generation squark searches can provide 
important information about the nature of SUSY parameter space and 
we hope that our work can be a starting guide to the experimental community 
to perform further analysis in this channel on the real data.

\section{Acknowledgements}
DG would like to acknowledge the hospitality of Prof. Palash B. Pal and the 
Theory Group of Saha Institute of Nuclear Physics where this work was finalized. 
DG and DS also thank Monoranjan Guchait for useful discussions and encouragement. 
DS acknowledges Christophe Grojean and the CERN theory division for the   
hospitality where part of the work was completed.

\section*{References}

\bibliography{sbottom.bib}{}

\providecommand{\href}[2]{#2}\begingroup\raggedright\begin{thebibliography}{10}

\bibitem{:2012gu}
{\bfseries CMS} Collaboration, S.~Chatrchyan {\em et~al.}, ``{Observation of a
  new boson at a mass of 125 GeV with the CMS experiment at the LHC},''
  \href{http://dx.doi.org/10.1016/j.physletb.2012.08.021}{{\em Phys.Lett.}
  {\bfseries B716} (2012) 30--61},
\href{http://arxiv.org/abs/1207.7235}{{\ttfamily arXiv:1207.7235 [hep-ex]}}.

\bibitem{:2012gk}
{\bfseries ATLAS} Collaboration, G.~Aad {\em et~al.}, ``{Observation of a new
  particle in the search for the Standard Model Higgs boson with the ATLAS
  detector at the LHC},''
  \href{http://dx.doi.org/10.1016/j.physletb.2012.08.020}{{\em Phys.Lett.}
  {\bfseries B716} (2012) 1--29},
\href{http://arxiv.org/abs/1207.7214}{{\ttfamily arXiv:1207.7214 [hep-ex]}}.

\bibitem{ATLAS-CONF-2012-103}
``Search for new phenomena using large jet multiplicities and missing
  transverse momentum with atlas in 5.8 fb$^{-1}$ of $\sqrt{s}=8$ tev
  proton-proton collisions,'' Tech. Rep. ATLAS-CONF-2012-103, CERN, Geneva,
  Aug, 2012.

\bibitem{ATLAS-CONF-2012-109}
``Search for squarks and gluinos with the atlas detector using final states
  with jets and missing transverse momentum and 5.8 fb$^{-1}$ of $\sqrt{s}$=8
  tev proton-proton collision data,'' Tech. Rep. ATLAS-CONF-2012-109, CERN,
  Geneva, Aug, 2012.

\bibitem{CMS-PAS-SUS-12-005}
``Search for supersymmetry with the razor variables at cms,''.

\bibitem{Baer:2011ab}
H.~Baer, V.~Barger, and A.~Mustafayev, ``{Implications of a 125 GeV Higgs
  scalar for LHC SUSY and neutralino dark matter searches},''
  \href{http://dx.doi.org/10.1103/PhysRevD.85.075010}{{\em Phys.Rev.}
  {\bfseries D85} (2012) 075010},
\href{http://arxiv.org/abs/1112.3017}{{\ttfamily arXiv:1112.3017 [hep-ph]}}.

\bibitem{Akula:2011aa}
S.~Akula, B.~Altunkaynak, D.~Feldman, P.~Nath, and G.~Peim, ``{Higgs Boson Mass
  Predictions in SUGRA Unification, Recent LHC-7 Results, and Dark Matter},''
  \href{http://dx.doi.org/10.1103/PhysRevD.85.075001}{{\em Phys.Rev.}
  {\bfseries D85} (2012) 075001},
\href{http://arxiv.org/abs/1112.3645}{{\ttfamily arXiv:1112.3645 [hep-ph]}}.

\bibitem{Feng:2011aa}
J.~L. Feng, K.~T. Matchev, and D.~Sanford, ``{Focus Point Supersymmetry
  Redux},'' \href{http://dx.doi.org/10.1103/PhysRevD.85.075007}{{\em Phys.Rev.}
  {\bfseries D85} (2012) 075007},
\href{http://arxiv.org/abs/1112.3021}{{\ttfamily arXiv:1112.3021 [hep-ph]}}.

\bibitem{Heinemeyer:2011aa}
S.~Heinemeyer, O.~Stal, and G.~Weiglein, ``{Interpreting the LHC Higgs Search
  Results in the MSSM},''
  \href{http://dx.doi.org/10.1016/j.physletb.2012.02.084}{{\em Phys.Lett.}
  {\bfseries B710} (2012) 201--206},
\href{http://arxiv.org/abs/1112.3026}{{\ttfamily arXiv:1112.3026 [hep-ph]}}.

\bibitem{Buchmueller:2011ab}
O.~Buchmueller, R.~Cavanaugh, A.~De~Roeck, M.~Dolan, J.~Ellis, {\em et~al.},
  ``{Higgs and Supersymmetry},''
  \href{http://dx.doi.org/10.1140/epjc/s10052-012-2020-3}{{\em Eur.Phys.J.}
  {\bfseries C72} (2012) 2020},
\href{http://arxiv.org/abs/1112.3564}{{\ttfamily arXiv:1112.3564 [hep-ph]}}.

\bibitem{Draper:2011aa}
P.~Draper, P.~Meade, M.~Reece, and D.~Shih, ``{Implications of a 125 GeV Higgs
  for the MSSM and Low-Scale SUSY Breaking},''
  \href{http://dx.doi.org/10.1103/PhysRevD.85.095007}{{\em Phys.Rev.}
  {\bfseries D85} (2012) 095007},
\href{http://arxiv.org/abs/1112.3068}{{\ttfamily arXiv:1112.3068 [hep-ph]}}.

\bibitem{Cao:2011sn}
J.~Cao, Z.~Heng, D.~Li, and J.~M. Yang, ``{Current experimental constraints on
  the lightest Higgs boson mass in the constrained MSSM},''
  \href{http://dx.doi.org/10.1016/j.physletb.2012.03.052}{{\em Phys.Lett.}
  {\bfseries B710} (2012) 665--670},
\href{http://arxiv.org/abs/1112.4391}{{\ttfamily arXiv:1112.4391 [hep-ph]}}.

\bibitem{Hall:2011aa}
L.~J. Hall, D.~Pinner, and J.~T. Ruderman, ``{A Natural SUSY Higgs Near 126
  GeV},'' \href{http://dx.doi.org/10.1007/JHEP04(2012)131}{{\em JHEP}
  {\bfseries 1204} (2012) 131},
\href{http://arxiv.org/abs/1112.2703}{{\ttfamily arXiv:1112.2703 [hep-ph]}}.

\bibitem{Ellis:2012aa}
J.~Ellis and K.~A. Olive, ``{Revisiting the Higgs Mass and Dark Matter in the
  CMSSM},'' \href{http://dx.doi.org/10.1140/epjc/s10052-012-2005-2}{{\em
  Eur.Phys.J.} {\bfseries C72} (2012) 2005},
\href{http://arxiv.org/abs/1202.3262}{{\ttfamily arXiv:1202.3262 [hep-ph]}}.

\bibitem{Cao:2012fz}
J.-J. Cao, Z.-X. Heng, J.~M. Yang, Y.-M. Zhang, and J.-Y. Zhu, ``{A SM-like
  Higgs near 125 GeV in low energy SUSY: a comparative study for MSSM and
  NMSSM},'' \href{http://dx.doi.org/10.1007/JHEP03(2012)086}{{\em JHEP}
  {\bfseries 1203} (2012) 086},
\href{http://arxiv.org/abs/1202.5821}{{\ttfamily arXiv:1202.5821 [hep-ph]}}.

\bibitem{Chang:2012gp}
C.-F. Chang, K.~Cheung, Y.-C. Lin, and T.-C. Yuan, ``{Mimicking the Standard
  Model Higgs Boson in UMSSM},''
  \href{http://dx.doi.org/10.1007/JHEP06(2012)128}{{\em JHEP} {\bfseries 1206}
  (2012) 128},
\href{http://arxiv.org/abs/1202.0054}{{\ttfamily arXiv:1202.0054 [hep-ph]}}.

\bibitem{Baer:2012uya}
H.~Baer, V.~Barger, and A.~Mustafayev, ``{Neutralino dark matter in
  mSUGRA/CMSSM with a 125 GeV light Higgs scalar},''
  \href{http://dx.doi.org/10.1007/JHEP05(2012)091}{{\em JHEP} {\bfseries 1205}
  (2012) 091},
\href{http://arxiv.org/abs/1202.4038}{{\ttfamily arXiv:1202.4038 [hep-ph]}}.

\bibitem{Ghosh:2012mc}
D.~Ghosh, M.~Guchait, and D.~Sengupta, ``{Higgs Signal in Chargino-Neutralino
  Production at the LHC},''
  \href{http://dx.doi.org/10.1140/epjc/s10052-012-2141-8}{{\em Eur.Phys.J.}
  {\bfseries C72} (2012) 2141},
\href{http://arxiv.org/abs/1202.4937}{{\ttfamily arXiv:1202.4937 [hep-ph]}}.

\bibitem{Maiani:2012ij}
L.~Maiani, A.~Polosa, and V.~Riquer, ``{Probing Minimal Supersymmetry at the
  LHC with the Higgs Boson Masses},''
  \href{http://dx.doi.org/10.1088/1367-2630/14/7/073029}{{\em New J.Phys.}
  {\bfseries 14} (2012) 073029},
\href{http://arxiv.org/abs/1202.5998}{{\ttfamily arXiv:1202.5998 [hep-ph]}}.

\bibitem{Cheng:2012np}
T.~Cheng, J.~Li, T.~Li, D.~V. Nanopoulos, and C.~Tong, ``{Electroweak
  Supersymmetry around the Electroweak Scale},''
\href{http://arxiv.org/abs/1202.6088}{{\ttfamily arXiv:1202.6088 [hep-ph]}}.

\bibitem{Jegerlehner:2012ju}
F.~Jegerlehner, ``{Implications of low and high energy measurements on SUSY
  models},'' {\em Frascati Phys.Ser.} {\bfseries 54} (2012) 42--51,
\href{http://arxiv.org/abs/1203.0806}{{\ttfamily arXiv:1203.0806 [hep-ph]}}.

\bibitem{Choudhury:2012tc}
A.~Choudhury and A.~Datta, ``{Many faces of low mass neutralino dark matter in
  the unconstrained MSSM, LHC data and new signals},''
  \href{http://dx.doi.org/10.1007/JHEP06(2012)006}{{\em JHEP} {\bfseries 1206}
  (2012) 006},
\href{http://arxiv.org/abs/1203.4106}{{\ttfamily arXiv:1203.4106 [hep-ph]}}.

\bibitem{Byakti:2012qk}
P.~Byakti and D.~Ghosh, ``{Magic Messengers in Gauge Mediation and signal for
  125 GeV boosted Higgs boson},''
\href{http://arxiv.org/abs/1204.0415}{{\ttfamily arXiv:1204.0415 [hep-ph]}}.

\bibitem{Brummer:2012ns}
F.~Brummer, S.~Kraml, and S.~Kulkarni, ``{Anatomy of maximal stop mixing in the
  MSSM},'' \href{http://dx.doi.org/10.1007/JHEP08(2012)089}{{\em JHEP}
  {\bfseries 1208} (2012) 089},
\href{http://arxiv.org/abs/1204.5977}{{\ttfamily arXiv:1204.5977 [hep-ph]}}.

\bibitem{Balazs:2012qc}
C.~Balazs, A.~Buckley, D.~Carter, B.~Farmer, and M.~White, ``{Should we still
  believe in constrained supersymmetry?},''
\href{http://arxiv.org/abs/1205.1568}{{\ttfamily arXiv:1205.1568 [hep-ph]}}.

\bibitem{Badziak:2012mm}
M.~Badziak, ``{Yukawa unification in SUSY SO(10) in light of the LHC Higgs
  data},'' \href{http://dx.doi.org/10.1142/S0217732312300200}{{\em
  Mod.Phys.Lett.} {\bfseries A27} (2012) 1230020},
\href{http://arxiv.org/abs/1205.6232}{{\ttfamily arXiv:1205.6232 [hep-ph]}}.

\bibitem{Feng:2012jf}
J.~L. Feng and D.~Sanford, ``{A Natural 125 GeV Higgs Boson in the MSSM from
  Focus Point Supersymmetry with A-Terms},''
  \href{http://dx.doi.org/10.1103/PhysRevD.86.055015}{{\em Phys.Rev.}
  {\bfseries D86} (2012) 055015},
\href{http://arxiv.org/abs/1205.2372}{{\ttfamily arXiv:1205.2372 [hep-ph]}}.

\bibitem{Ghosh:2012dh}
D.~Ghosh, M.~Guchait, S.~Raychaudhuri, and D.~Sengupta, ``{How Constrained is
  the cMSSM?},'' \href{http://dx.doi.org/10.1103/PhysRevD.86.055007}{{\em
  Phys.Rev.} {\bfseries D86} (2012) 055007},
\href{http://arxiv.org/abs/1205.2283}{{\ttfamily arXiv:1205.2283 [hep-ph]}}.

\bibitem{Dudas:2012hx}
E.~Dudas, Y.~Mambrini, A.~Mustafayev, and K.~A. Olive, ``{Relating the CMSSM
  and SUGRA Models with GUT Scale and Super-GUT Scale Supersymmetry
  Breaking},'' \href{http://dx.doi.org/10.1140/epjc/s10052-012-2138-3}{{\em
  Eur.Phys.J.} {\bfseries C72} (2012) 2138},
\href{http://arxiv.org/abs/1205.5988}{{\ttfamily arXiv:1205.5988 [hep-ph]}}.

\bibitem{Fowlie:2012im}
A.~Fowlie, M.~Kazana, K.~Kowalska, S.~Munir, L.~Roszkowski, {\em et~al.},
  ``{The CMSSM Favoring New Territories: The Impact of New LHC Limits and a 125
  GeV Higgs},'' \href{http://dx.doi.org/10.1103/PhysRevD.86.075010}{{\em
  Phys.Rev.} {\bfseries D86} (2012) 075010},
\href{http://arxiv.org/abs/1206.0264}{{\ttfamily arXiv:1206.0264 [hep-ph]}}.

\bibitem{Athron:2012sq}
P.~Athron, S.~King, D.~Miller, S.~Moretti, and R.~Nevzorov, ``{Constrained
  Exceptional Supersymmetric Standard Model with a Higgs Near 125 GeV},''
  \href{http://dx.doi.org/10.1103/PhysRevD.86.095003}{{\em Phys.Rev.}
  {\bfseries D86} (2012) 095003},
\href{http://arxiv.org/abs/1206.5028}{{\ttfamily arXiv:1206.5028 [hep-ph]}}.

\bibitem{Chatterjee:2012qt}
R.~M. Chatterjee, M.~Guchait, and D.~Sengupta, ``{Probing Supersymmetry using
  Event Shape variables at 8 TeV LHC},''
  \href{http://dx.doi.org/10.1103/PhysRevD.86.075014}{{\em Phys.Rev.}
  {\bfseries D86} (2012) 075014},
\href{http://arxiv.org/abs/1206.5770}{{\ttfamily arXiv:1206.5770 [hep-ph]}}.

\bibitem{CahillRowley:2012rv}
M.~W. Cahill-Rowley, J.~L. Hewett, A.~Ismail, and T.~G. Rizzo, ``{The Higgs
  Sector and Fine-Tuning in the pMSSM},''
  \href{http://dx.doi.org/10.1103/PhysRevD.86.075015}{{\em Phys.Rev.}
  {\bfseries D86} (2012) 075015},
\href{http://arxiv.org/abs/1206.5800}{{\ttfamily arXiv:1206.5800 [hep-ph]}}.

\bibitem{Akula:2012kk}
S.~Akula, P.~Nath, and G.~Peim, ``{Implications of the Higgs Boson Discovery
  for mSUGRA},'' \href{http://dx.doi.org/10.1016/j.physletb.2012.09.007}{{\em
  Phys.Lett.} {\bfseries B717} (2012) 188--192},
\href{http://arxiv.org/abs/1207.1839}{{\ttfamily arXiv:1207.1839 [hep-ph]}}.

\bibitem{Cao:2012yn}
J.~Cao, Z.~Heng, J.~M. Yang, and J.~Zhu, ``{Status of low energy SUSY models
  confronted with the LHC 125 GeV Higgs data},''
  \href{http://dx.doi.org/10.1007/JHEP10(2012)079}{{\em JHEP} {\bfseries 1210}
  (2012) 079},
\href{http://arxiv.org/abs/1207.3698}{{\ttfamily arXiv:1207.3698 [hep-ph]}}.

\bibitem{Arbey:2012dq}
A.~Arbey, M.~Battaglia, A.~Djouadi, and F.~Mahmoudi, ``{The Higgs sector of the
  phenomenological MSSM in the light of the Higgs boson discovery},''
  \href{http://dx.doi.org/10.1007/JHEP09(2012)107}{{\em JHEP} {\bfseries 1209}
  (2012) 107},
\href{http://arxiv.org/abs/1207.1348}{{\ttfamily arXiv:1207.1348 [hep-ph]}}.

\bibitem{Nath:2012fa}
P.~Nath, ``{SUGRA Grand Unification, LHC and Dark Matter},''
\href{http://arxiv.org/abs/1207.5501}{{\ttfamily arXiv:1207.5501 [hep-ph]}}.

\bibitem{Desai:2011th}
N.~Desai and B.~Mukhopadhyaya, ``{Constraints on supersymmetry with light third
  family from LHC data},''
  \href{http://dx.doi.org/10.1007/JHEP05(2012)057}{{\em JHEP} {\bfseries 1205}
  (2012) 057},
\href{http://arxiv.org/abs/1111.2830}{{\ttfamily arXiv:1111.2830 [hep-ph]}}.

\bibitem{He:2011tp}
B.~He, T.~Li, and Q.~Shafi, ``{Impact of LHC Searches on NLSP Top Squark and
  Gluino Mass},'' \href{http://dx.doi.org/10.1007/JHEP05(2012)148}{{\em JHEP}
  {\bfseries 1205} (2012) 148},
\href{http://arxiv.org/abs/1112.4461}{{\ttfamily arXiv:1112.4461 [hep-ph]}}.

\bibitem{Drees:2012dd}
M.~Drees, M.~Hanussek, and J.~S. Kim, ``{Light Stop Searches at the LHC with
  Monojet Events},'' \href{http://dx.doi.org/10.1103/PhysRevD.86.035024}{{\em
  Phys.Rev.} {\bfseries D86} (2012) 035024},
\href{http://arxiv.org/abs/1201.5714}{{\ttfamily arXiv:1201.5714 [hep-ph]}}.

\bibitem{Berger:2012ec}
J.~Berger, J.~Hubisz, and M.~Perelstein, ``{A Fermionic Top Partner:
  Naturalness and the LHC},''
  \href{http://dx.doi.org/10.1007/JHEP07(2012)016}{{\em JHEP} {\bfseries 1207}
  (2012) 016},
\href{http://arxiv.org/abs/1205.0013}{{\ttfamily arXiv:1205.0013 [hep-ph]}}.

\bibitem{Plehn:2012pr}
T.~Plehn, M.~Spannowsky, and M.~Takeuchi, ``{Stop searches in 2012},''
  \href{http://dx.doi.org/10.1007/JHEP08(2012)091}{{\em JHEP} {\bfseries 1208}
  (2012) 091},
\href{http://arxiv.org/abs/1205.2696}{{\ttfamily arXiv:1205.2696 [hep-ph]}}.

\bibitem{Han:2012fw}
Z.~Han, A.~Katz, D.~Krohn, and M.~Reece, ``{(Light) Stop Signs},''
  \href{http://dx.doi.org/10.1007/JHEP08(2012)083}{{\em JHEP} {\bfseries 1208}
  (2012) 083},
\href{http://arxiv.org/abs/1205.5808}{{\ttfamily arXiv:1205.5808 [hep-ph]}}.

\bibitem{Barger:2012hr}
V.~Barger, P.~Huang, M.~Ishida, and W.-Y. Keung, ``{Scalar-Top Masses from SUSY
  Loops with 125 GeV mh and Precise Mw},''
\href{http://arxiv.org/abs/1206.1777}{{\ttfamily arXiv:1206.1777 [hep-ph]}}.

\bibitem{Choudhury:2012kn}
A.~Choudhury and A.~Datta, ``{New limits on top squark NLSP from LHC 4.7
  $fb^{-1}$ data},'' \href{http://dx.doi.org/10.1142/S021773231250188X}{{\em
  Mod.Phys.Lett.} {\bfseries A27} (2012) 1250188},
\href{http://arxiv.org/abs/1207.1846}{{\ttfamily arXiv:1207.1846 [hep-ph]}}.

\bibitem{Cao:2012rz}
J.~Cao, C.~Han, L.~Wu, J.~M. Yang, and Y.~Zhang, ``{Probing Natural SUSY from
  Stop Pair Production at the LHC},''
\href{http://arxiv.org/abs/1206.3865}{{\ttfamily arXiv:1206.3865 [hep-ph]}}.

\bibitem{Ghosh:2012ud}
K.~Ghosh, K.~Huitu, J.~Laamanen, L.~Leinonen, K.~Huitu, {\em et~al.}, ``{Top
  jets as a probe of degenerate stop-NLSP LSP scenario in the framework of
  cMSSM},''
\href{http://arxiv.org/abs/1207.2429}{{\ttfamily arXiv:1207.2429 [hep-ph]}}.

\bibitem{Dutta:2012kx}
B.~Dutta, T.~Kamon, N.~Kolev, K.~Sinha, and K.~Wang, ``{Searching for Top
  Squarks at the LHC in Fully Hadronic Final State},''
  \href{http://dx.doi.org/10.1103/PhysRevD.86.075004}{{\em Phys.Rev.}
  {\bfseries D86} (2012) 075004},
\href{http://arxiv.org/abs/1207.1873}{{\ttfamily arXiv:1207.1873 [hep-ph]}}.

\bibitem{Chen:2012uw}
C.-Y. Chen, A.~Freitas, T.~Han, and K.~S. Lee, ``{New Physics from the Top at
  the LHC},''
\href{http://arxiv.org/abs/1207.4794}{{\ttfamily arXiv:1207.4794 [hep-ph]}}.

\bibitem{Bartl:1997yi}
A.~Bartl, H.~Eberl, S.~Kraml, W.~Majerotto, W.~Porod, {\em et~al.}, ``{Search
  of stop, sbottom, tau sneutrino, and stau at an e+ e- linear collider with
  S**(1/2) = 0.5-TeV - 2-TeV},''
  \href{http://dx.doi.org/10.1007/s002880050577}{{\em Z.Phys.} {\bfseries C76}
  (1997) 549--560},
\href{http://arxiv.org/abs/hep-ph/9701336}{{\ttfamily arXiv:hep-ph/9701336
  [hep-ph]}}.

\bibitem{Hisano:2003qu}
J.~Hisano, K.~Kawagoe, and M.~M. Nojiri, ``{A Detailed study of the gluino
  decay into the third generation squarks at the CERN LHC},''
  \href{http://dx.doi.org/10.1103/PhysRevD.68.035007}{{\em Phys.Rev.}
  {\bfseries D68} (2003) 035007},
\href{http://arxiv.org/abs/hep-ph/0304214}{{\ttfamily arXiv:hep-ph/0304214
  [hep-ph]}}.

\bibitem{Drees:2000he}
M.~Drees, Y.~G. Kim, M.~M. Nojiri, D.~Toya, K.~Hasuko, {\em et~al.},
  ``{Scrutinizing LSP dark matter at the CERN LHC},''
  \href{http://dx.doi.org/10.1103/PhysRevD.63.035008}{{\em Phys.Rev.}
  {\bfseries D63} (2001) 035008},
\href{http://arxiv.org/abs/hep-ph/0007202}{{\ttfamily arXiv:hep-ph/0007202
  [hep-ph]}}.

\bibitem{Hisano:2002xq}
J.~Hisano, K.~Kawagoe, R.~Kitano, and M.~M. Nojiri, ``{Scenery from the top:
  Study of the third generation squarks at CERN LHC},''
  \href{http://dx.doi.org/10.1103/PhysRevD.66.115004}{{\em Phys.Rev.}
  {\bfseries D66} (2002) 115004},
\href{http://arxiv.org/abs/hep-ph/0204078}{{\ttfamily arXiv:hep-ph/0204078
  [hep-ph]}}.

\bibitem{Alves:2007xt}
A.~Alves and O.~Eboli, ``{Unravelling the sbottom spin at the CERN LHC},''
  \href{http://dx.doi.org/10.1103/PhysRevD.75.115013}{{\em Phys.Rev.}
  {\bfseries D75} (2007) 115013},
\href{http://arxiv.org/abs/0704.0254}{{\ttfamily arXiv:0704.0254 [hep-ph]}}.

\bibitem{Belyaev:2009wf}
A.~Belyaev, T.~Lastovicka, A.~Nomerotski, and G.~Lastovicka-Medin,
  ``{Discovering Bottom Squark Co-annihilation at ILC},''
  \href{http://dx.doi.org/10.1103/PhysRevD.81.035011}{{\em Phys.Rev.}
  {\bfseries D81} (2010) 035011},
\href{http://arxiv.org/abs/0912.2411}{{\ttfamily arXiv:0912.2411 [hep-ph]}}.

\bibitem{AdeelAjaib:2011ec}
M.~Adeel~Ajaib, T.~Li, and Q.~Shafi, ``{Searching for NLSP Sbottom at the
  LHC},'' \href{http://dx.doi.org/10.1016/j.physletb.2011.05.059}{{\em
  Phys.Lett.} {\bfseries B701} (2011) 255--259},
\href{http://arxiv.org/abs/1104.0251}{{\ttfamily arXiv:1104.0251 [hep-ph]}}.

\bibitem{Lee:2012sy}
H.~M. Lee, V.~Sanz, and M.~Trott, ``{Hitting sbottom in natural SUSY},''
  \href{http://dx.doi.org/10.1007/JHEP05(2012)139}{{\em JHEP} {\bfseries 1205}
  (2012) 139},
\href{http://arxiv.org/abs/1204.0802}{{\ttfamily arXiv:1204.0802 [hep-ph]}}.

\bibitem{Alvarez:2012wf}
E.~Alvarez and Y.~Bai, ``{Reach the Bottom Line of the Sbottom Search},''
  \href{http://dx.doi.org/10.1007/JHEP08(2012)003}{{\em JHEP} {\bfseries 1208}
  (2012) 003},
\href{http://arxiv.org/abs/1204.5182}{{\ttfamily arXiv:1204.5182 [hep-ph]}}.

\bibitem{Ajaib:2012eb}
M.~A. Ajaib, I.~Gogoladze, and Q.~Shafi, ``{Higgs Boson Production and Decay:
  Effects from Light Third Generation and Vectorlike Matter},''
\href{http://arxiv.org/abs/1207.7068}{{\ttfamily arXiv:1207.7068 [hep-ph]}}.

\bibitem{CMS:2012nxa}
{\bfseries CMS} Collaboration, CMS-PAS-SUS-11-022,
``{Search for supersymmetery in final states with missing transverse momentum
  and 0, 1, 2, or $\ge 3$ b jets with CMS},''.

\bibitem{Chatrchyan:2012sa}
{\bfseries CMS} Collaboration, S.~Chatrchyan {\em et~al.}, ``{Search for new
  physics in events with same-sign dileptons and b-tagged jets in pp collisions
  at sqrt(s) = 7 TeV},'' \href{http://dx.doi.org/10.1007/JHEP08(2012)110}{{\em
  JHEP} {\bfseries 1208} (2012) 110},
\href{http://arxiv.org/abs/1205.3933}{{\ttfamily arXiv:1205.3933 [hep-ex]}}.

\bibitem{Aad:2012pq}
{\bfseries ATLAS} Collaboration, G.~Aad {\em et~al.}, ``{Search for top and
  bottom squarks from gluino pair production in final states with missing
  transverse energy and at least three b-jets with the ATLAS detector},''
\href{http://arxiv.org/abs/1207.4686}{{\ttfamily arXiv:1207.4686 [hep-ex]}}.

\bibitem{Martin:1997ns}
S.~P. Martin, ``{A Supersymmetry primer},''
\href{http://arxiv.org/abs/hep-ph/9709356}{{\ttfamily arXiv:hep-ph/9709356
  [hep-ph]}}.

\bibitem{Djouadi:1998di}
{\bfseries MSSM Working Group} Collaboration, A.~Djouadi {\em et~al.}, ``{The
  Minimal supersymmetric standard model: Group summary report},''
\href{http://arxiv.org/abs/hep-ph/9901246}{{\ttfamily arXiv:hep-ph/9901246
  [hep-ph]}}.

\bibitem{PhysRevD.86.010001}
{\bfseries Particle Data Group} Collaboration, J.~Beringer {\em et~al.},
  ``Review of particle physics,''
  \href{http://dx.doi.org/10.1103/PhysRevD.86.010001}{{\em Phys. Rev. D}
  {\bfseries 86} (Jul, 2012) 010001}.

\bibitem{Carena:2012gp}
M.~Carena, S.~Gori, N.~R. Shah, C.~E. Wagner, and L.-T. Wang, ``{Light Stau
  Phenomenology and the Higgs $\gamma\gamma$ Rate},''
  \href{http://dx.doi.org/10.1007/JHEP07(2012)175}{{\em JHEP} {\bfseries 1207}
  (2012) 175},
\href{http://arxiv.org/abs/1205.5842}{{\ttfamily arXiv:1205.5842 [hep-ph]}}.

\bibitem{Beenakker:1996ed}
W.~Beenakker, R.~Hopker, and M.~Spira, ``{PROSPINO: A Program for the
  production of supersymmetric particles in next-to-leading order QCD},''
\href{http://arxiv.org/abs/hep-ph/9611232}{{\ttfamily arXiv:hep-ph/9611232
  [hep-ph]}}.

\bibitem{Djouadi:2002ze}
A.~Djouadi, J.-L. Kneur, and G.~Moultaka, ``{SuSpect: A Fortran code for the
  supersymmetric and Higgs particle spectrum in the MSSM},''
  \href{http://dx.doi.org/10.1016/j.cpc.2006.11.009}{{\em Comput.Phys.Commun.}
  {\bfseries 176} (2007) 426--455},
\href{http://arxiv.org/abs/hep-ph/0211331}{{\ttfamily arXiv:hep-ph/0211331
  [hep-ph]}}.

\bibitem{Sjostrand:2006za}
T.~Sjostrand, S.~Mrenna, and P.~Z. Skands, ``{PYTHIA 6.4 Physics and Manual},''
  \href{http://dx.doi.org/10.1088/1126-6708/2006/05/026}{{\em JHEP} {\bfseries
  0605} (2006) 026},
\href{http://arxiv.org/abs/hep-ph/0603175}{{\ttfamily arXiv:hep-ph/0603175
  [hep-ph]}}.

\bibitem{Cacciari:2011ma}
M.~Cacciari, G.~P. Salam, and G.~Soyez, ``{FastJet User Manual},''
  \href{http://dx.doi.org/10.1140/epjc/s10052-012-1896-2}{{\em Eur.Phys.J.}
  {\bfseries C72} (2012) 1896},
\href{http://arxiv.org/abs/1111.6097}{{\ttfamily arXiv:1111.6097 [hep-ph]}}.

\bibitem{Cacciari:2008gp}
M.~Cacciari, G.~P. Salam, and G.~Soyez, ``{The Anti-k(t) jet clustering
  algorithm},'' \href{http://dx.doi.org/10.1088/1126-6708/2008/04/063}{{\em
  JHEP} {\bfseries 0804} (2008) 063},
\href{http://arxiv.org/abs/0802.1189}{{\ttfamily arXiv:0802.1189 [hep-ph]}}.

\bibitem{Lai:1999wy}
{\bfseries CTEQ} Collaboration, H.~Lai {\em et~al.}, ``{Global QCD analysis of
  parton structure of the nucleon: CTEQ5 parton distributions},''
  \href{http://dx.doi.org/10.1007/s100529900196}{{\em Eur.Phys.J.} {\bfseries
  C12} (2000) 375--392},
\href{http://arxiv.org/abs/hep-ph/9903282}{{\ttfamily arXiv:hep-ph/9903282
  [hep-ph]}}.

\bibitem{Bourilkov:2006cj}
D.~Bourilkov, R.~C. Group, and M.~R. Whalley, ``{LHAPDF: PDF use from the
  Tevatron to the LHC},''
\href{http://arxiv.org/abs/hep-ph/0605240}{{\ttfamily arXiv:hep-ph/0605240
  [hep-ph]}}.

\bibitem{Kling:2012up}
F.~Kling, T.~Plehn, and M.~Takeuchi, ``{Tagging single Tops},''
\href{http://arxiv.org/abs/1207.4787}{{\ttfamily arXiv:1207.4787 [hep-ph]}}.

\bibitem{CMS-PAS-BTV-11-001}
{\bfseries CMS} Collaboration, CMS-PAS-BTV-11-001, ``{Performance of the b-jet
  identification in CMS},''.

\end{thebibliography}\endgroup
\bibliographystyle{utphys.bst}

\end{document}